\documentstyle[12pt]{article}
\if@twoside  m
\else
\fi \topmargin 5mm \headheight 0mm \headsep 0mm \textheight
225truemm \textwidth 150truemm
\newcommand{\ie}{{\em i.e.}}

\newcommand{\cf}{{\em cf.\ }}
\newcommand{\rhs}{{r.h.s.\ }}

\newcommand{\R}{I\!\!R}
\newcommand{\C}{C\!\!\!\rule[.5pt]{.7pt}{6.5pt}\:\:}
\newcommand{\BB}{{\cal B}}
\newcommand{\OO}{{\cal O}}
\newcommand{\aleq}{\stackrel{<}{\scriptstyle\sim}}
\begin{document}

\title{Anomalous electron trapping by localized \\ magnetic
fields }
\author{F. Bentosela,$^{a,b}$ R.M. Cavalcanti,$^{c}$
P. Exner,$^{d,e}$ V.A. Zagrebnov$^{a,b}$}
\date{}
\maketitle
\begin{quote}
{\small \em a) Centre de Physique Th\'{e}orique, C.N.R.S., 13288
Marseille Luminy \\ b) Universit\'{e} de la Mediterran\'{e}e
(Aix--Marseille II), 13288 Marseille \\ \phantom{e)x} Luminy,
France
\\ c) Institute for Theoretical Physics, University of California,
Santa Barbara, \phantom{e)x} CA 93106-4030, USA \\ d) Nuclear
Physics Institute, Academy of Sciences, 25068 \v Re\v z \\ e)
Doppler Institute, Czech Technical University, B\v{r}ehov{\'a} 7,
11519 Prague, \phantom{e)x} Czech Republic \\ \rm
\phantom{e)x}bentosela@cpt.univ-mrs.fr, ricardomc@hotmail.com, \\
\phantom{e)x}exner@ujf.cas.cz, zagrebnov@cpt.univ-mrs.fr}
\vspace{8mm}

\noindent {\small We consider an electron with an anomalous
magnetic moment $g>2$ confined to a plane and interacting with a
nonzero magnetic field $B$ perpendicular to the plane.
We show that if $B$ has compact
support and the magnetic flux in the natural units is $F\ge 0$,
the corresponding Pauli Hamiltonian has at least $1+[F]$ bound
states, without making any assumptions about the field profile.
Furthermore, in the zero-flux case there is a pair of bound states
with opposite spin orientations. Using a Birman-Schwinger
technique, we extend the last claim to a weak rotationally
symmetric field with $B(r) = \OO(r^{-2-\delta})$ correcting thus a
recent result. Finally, we show that under mild regularity
assumptions the existence can be proved for non-symmetric fields
with tails as well.}
\end{quote}

%%%%%%%%%%%%%%%%%%%%%%%%%%%%%%%%%%%%%%%%%%%%%%%%%%%%%%%%%%%%%%%%%%%

\section{Introduction}

Interaction of electrons with a localized magnetic field has been
a subject of interest for a long time. It has been observed
recently that a magnetic flux tube can bind charged particles with
anomalous magnetic moment $g>2$. An example of such a particle
is the electron, which has $\,g=2.0023$.

The effect was observed first in simple examples \cite{BV,CFC,Mo}
such as a cylindrical tube with a field which is either homogeneous
or supported by the tube surface. The same behaviour has been then
demonstrated for any rotationally invariant field $B(r)$ with
compact support and which does not change sign \cite{CC}. In the
next step the symmetry condition was removed and the
positivity requirement weakened \cite{BEZ2}. The main aim of the
present paper is to complete this process by showing that bound
states exist for any (nontrivial) compactly supported field and
their number is controlled by the number of flux quanta: the
corresponding Pauli Hamiltonian will be shown to have at least
$1+[F]$ negative eigenvalues, where $F$ is the value of
the flux through the tube in natural units.

This improvement is made possible by a pair of new tools. First of
all, the supersymmetry properties of Pauli operators allow us to
show that the matrix element of the field appearing in the
sufficient condition of \cite{BEZ2} has in fact a definite sign.
This trick will be combined with a more sophisticated variational
estimate which enables us to treat the integer-flux situation at
the same footing as the other cases. In particular, we will be
able to demonstrate in this way that for a nonzero $B$ a bound
state due to an excess magnetic moment exists even if the flux is
zero. More than that, an analogous argument shows that in this
situation the field binds electrons with {\em both} spin
orientations.

While the proof of the last result requires a compact support and does not
cover fields with tails extending to infinity, it raises a
question about a claim made in a recent paper by some of us
\cite{BEZ1}. It was said there that a system with a particular
rotationally symmetric field induced by an electric current vortex
has no bound states for weak currents. This is not correct: the
statement is true for higher partial waves only, while the s-wave
part has in reality a nontrivial spectrum for any nonzero current.

The error is subtle and --- as we hope --- instructive: it
illustrates well the fine nature of weakly bound states of
Schr\"odinger operators in one and two dimensions. The point is
that caution is needed when the coupling is switched off {\em
nonlinearly\/}: the case in question represents an example of
a two-dimensional Schr\"odinger operator with a potential which
has a {\em positive mean} for any nonzero coupling constant while
still having a bound state.

To set things straight we discuss in the last three sections the
weak field, zero-flux case in detail by performing
the corresponding Birman-Schwinger analysis to second order.
For centrally symmetric fields it yields $g>2$ as a sufficient
condition for the existence of bound states, and
provides an asymptotic formula for the bound-state energy. We
also show that adding some regularity assumptions one can prove in
this way the existence of weakly bound states for
non-symmetric fields with tails as well.

%%%%%%%%%%%%%%%%%%%%%%%%%%%%%%%%%%%%%%%%%%%%%%%%%%%%%%%%%%%%%%%%%%%%%

\setcounter{equation}{0}
\section{Preliminaries}

As we have said we consider a particle of spin $1/2$ living in a
plane and subject to a non-homogeneous magnetic field $B$
perpendicular to the plane. Here and in the next section we
suppose that $B$ has support in a compact region $\Sigma$ of
$\R^2$; later we shall replace this by a suitable decay
requirement. No hypotheses are made here about the field profile;
we assume just its integrability, $B\in L^1(\Sigma)$. The
corresponding vector potential $A= (A_1,A_2)$ lies in the plane
and $B= \partial_1A_2-\partial_2A_1$. Throughout the paper we
employ natural units, $2m= \hbar= c= e= 1$. \vspace{2mm}

\noindent {\bf Remark.} The assumptions do not include the singular
field profile $B(x)= 2\pi F\,\delta(x)$ (a magnetic string).
Although it can be regarded as a squeezing limit of $L^1$
fields, the procedure is non-trivial: as pointed out in
\cite{BV} one has to perform at the same time the non-physical
limit $g\to 2$ to preserve the existence of bound states in
analogy with the coupling constant renormalization for the usual
two-dimensional $\delta$ interaction \cite{AGHH}. We will not
discuss this case here. \vspace{2mm}

The particle dynamics is described by the two--dimensional
Pauli Hamiltonian  which we write in the standard form \cite{Th},
\begin{equation} \label{Pauli}
H_P^{(\pm)}(A)=\left( -i\nabla-A(x)\right)^2\pm{g\over 2}\,
B(x) = D^*D+{1\over 2}\,(2\pm g) B(x)\,,
\end{equation}
where $D:= (p_1\!-A_1)+ i(p_2\!-A_2)$ and the two signs
correspond to the two possible spin orientations. The quantity
\begin{equation} \label{flux}
F:= {1\over 2\pi}\, \int_{\Sigma} B(x)\, d^2x
\end{equation}
is the total flux measured in the natural units $(2\pi)^{-1}$,
or the number of flux quanta through $\Sigma$. We
assume conventionally that $F\ge 0$, {\ie}, if the mean field
is nonzero it points up. In such a case we will be interested
primarily in the operator $H_P^{(-)}(A)$ which describes an
electron with its magnetic moment parallel to the flux.

Next we have to recall a classical result of Aharonov and Casher
\cite{AC,Th} which will be a basic ingredient of our argument in
the next section. It says that if $F= N+\varepsilon$,
$\varepsilon\in (0,1]$ for a positive integer $N$, the
operator $H_P^{(-)}(A)$ with nonanomalous moment, $g=2$,
has $N$ zero energy eigenvalues. In the gauge
$A_1= -\partial_2 \phi$, $A_2=\partial_1 \phi$, where
\begin{equation} \label{phi}
\phi(x):= {1\over 2\pi} \int_{\Sigma} B(y)\,\ln|x-y|\,
d^2y \,,
\end{equation}
the corresponding eigenfunctions are given explicitly by
\begin{equation} \label{AC ef's}
\chi_j(x)= e^{-\phi(x)}\, (x_1+ix_2)^j, \quad j=0,1,\dots,
N-1\,.
\end{equation}
It is easy to check that $D\chi_j = 0$ for any nonnegative integer
$j$, but only those functions listed in (\ref{AC ef's}) are
square-integrable; this follows from the fact that $\chi_j(x)=
\OO(|x|^{-F+j})$ as $|x|\to\infty$ --- \cf \cite{AC},
\cite[Sec.7.2]{Th}. However, the functions $\chi_j$ with $j=[F]$
and $j=[F]-1$ (the latter in the case $F$ is a positive integer;
as usual, the symbol $[\cdot]$ denotes the integer part) are zero
energy resonances, since they solve the equation
$H_P^{(-)}(A)\chi_j=0$ and do not grow at large distances.

%%%%%%%%%%%%%%%%%%%%%%%%%%%%%%%%%%%%%%%%%%%%%%%%%%%%%%%%%%%%%%%%%%%%%%%

\setcounter{equation}{0}
\section{Flux tubes}

Now we are in position to state our main result about the
existence and number of bound states of the operator
(\ref{Pauli}). \vspace{3mm}

\noindent {\bf Theorem 1.} If $B\in L^1$ is nonzero and
compactly supported, the operator $H_P^{(-)}(A)$ has for
$g>2$ at least $1+[F]$ negative eigenvalues. Moreover, if
$F=0$ then $H_P^{(+)}(A)$ also has a bound state.
\vspace{2mm}

\noindent {\em Proof:} By the minimax principle, it is sufficient
for the first claim to find a subspace of dimension $1+[F]$ on
which the quadratic form
$$ (\psi,H_P^{(-)}(A)\psi)= \int_{\R^2} |(D\psi)(x)|^2\, d^2x
-{1\over 2}\,(g-2) \int_{\R^2} B(x) |\psi(x)|^2\, d^2x $$
is negative. To construct appropriate trial functions
$\psi_{\alpha}$ we employ the above mentioned zero--energy
solutions; specifically, we choose
\begin{equation} \label{psi}
\psi_{\alpha}(x) = \sum_{j=0}^{[F]} \alpha_j \left(
f_{R,\kappa}(r)\chi_j(x) +\varepsilon h_j(x) \right)\,,
\end{equation}
where $h_j\in C_0^{2}(\Sigma)$ will be specified later and
$f_{R,\kappa}: \R_+\to \R$ is a suitable function such that
$f_{R,\kappa}(r)=1$ for $r:=|x|\le R$, with $R$ chosen in such a
way that $\Sigma$ is a subset of $\BB_R:= \{x:\; |x| \le R \}$.
Clearly it is sufficient to consider coefficient vectors
$\,\alpha\in\C^{1+[F]}$ with $\,|\alpha|=1\,$.

It is straightforward to compute the value of the energy
form; with a later purpose on mind we write it as
\begin{eqnarray}\label{energy form}
(\psi_{\alpha},(D^*D+\mu B)\psi_{\alpha}) &\!=\!&
\sum_{j,k=0}^{[F]} \bar\alpha_j \alpha_k \left\{ \int_{\R^2}
\left|f'_{R,\kappa}(r)\right|^2 (\bar\chi_j \chi_k)(x)\, d^2x
\right. \\
&\!+\!& \varepsilon^2\, \int_{\Sigma} (\overline{Dh_j})(x)
(Dh_k)(x)\, d^2x + \mu\left[ \int_{\Sigma} (B \bar\chi_j
\chi_k)(x)\,d^2x \right. \nonumber \\
&\!+\!& \varepsilon\,
\int_{\Sigma} ((\bar h_j\chi_k + \bar\chi_j h_k)B)(x)\, d^2x +
\left. \left. \varepsilon^2 \int_{\Sigma} (B \bar h_j h_k)(x)\,
d^2x \right] \right\} \nonumber
\end{eqnarray}
with $\mu= -{1\over 2}\,(g\!-\!2)$. We have employed here the
property $D\chi_j=0$ of the AC functions and the fact that $h_j$
and $f'_{R,\kappa}$ have by assumption disjoint supports: we have
$D \Sigma_j \alpha_j f_{R,\kappa}\chi_j = 0$ inside $\Sigma$ so
$D\psi_{\alpha} = \varepsilon \Sigma_j \alpha_j Dh_j$ there, while
outside $\Sigma$ we have $h_j=0$, so $D\psi_{\alpha} =
Df_{R,\kappa}\Sigma_j \alpha_j \chi_j = \Sigma_j \alpha_j \chi_j
(p_1+ip_2) f_{R,\kappa} = \chi_{\alpha} (-ix_1+x_2) r^{-1}
f'_{R,\kappa}$.

As a warm-up, suppose first that $F-j>1$ holds for all nonzero
coefficients $\alpha_j$. Then the corresponding $\chi_j\in L^2$
and we can use the simplest choice $f_{R,\kappa}=1$ and
$\varepsilon=0$ in (\ref{psi}) obtaining
\begin{equation} \label{L^2 bound}
(\psi_{\alpha},H_P^{(-)}(A)\psi_{\alpha})=- {1\over 2}\,(g-2)
\int_{\Sigma} B(x)|\psi_{\alpha}(x)|^2\, d^2x\,. \end{equation}
Suppose that $(\psi_{\alpha},B\psi_{\alpha})\le 0$. Since
$D^*D\psi_{\alpha}=0$, this would imply the inequality
$(\psi_{\alpha},(D^*D+2B)\psi_{\alpha})\le 0$, but the operator in
parentheses equals $DD^*$ giving thus $\|D^*\psi_{\alpha}\|^2 \le
0$. This is possible only if $D^*\psi_{\alpha}=0$, which is false,
because otherwise we would have $2B\psi_{\alpha}=
(DD^*-D^*D)\psi_{\alpha} =0$ or $B(x)\psi_{\alpha}=0$ almost
everywhere. Since $\psi_{\alpha}$ is a product of a positive
function $e^{-\phi(x)}$ and a polynomial in $x_1+ix_2$, it has at
most $[F]-1 $ zeros --- recall that we are assuming $j<F-1$ ---
and we arrive at $B(x)=0$ a.e.\ which contradicts the assumption.
Consequently, the \rhs of (\ref{L^2 bound}) is negative for $g>2$.

If the linear combination includes $\alpha_j$ with $0\le F-j\le
1$, the situation is more complicated. Since the corresponding AC
functions are no longer $L^2$, we have to modify the trial
function at large distances, but gently enough to make the
positive energy contribution from the tails small. We achieve that
by choosing
\begin{equation} \label{tails}
f_{R,\kappa}(r) := \min\left\{ 1,\, {K_0(\kappa r)\over
K_0(\kappa R)} \right\},
\end{equation}
where $K_0$ is the Macdonald function and the parameter
$\kappa$ will be specified later. Since $K_0$ is
strictly decreasing, the corresponding $\psi_{\alpha}$ will not be
smooth at $r=R$ but it remains continuous, hence it
is an admissible trial function. To estimate the first term at the
\rhs of (\ref{energy form}), let us compute
\begin{eqnarray*} K_0(\kappa R)^2 \int_{\R^2}
|f'_{R,\kappa}(r)|^2\,d^2x &\!=\!& 2\pi \int_{\kappa R}^{\infty}
K_1(t)^2\,t\,dt \\
&\!=\!& \pi \left[\kappa^2R^2 K'_1(\kappa R)^2 -
\left(\kappa^2 R^2\!+1\right) K_1(\kappa R)^2 \,\right],
\end{eqnarray*}
\cf \cite[9.6.26]{AS}, \cite[1.12.3.2]{PBM}. Using
$-K'_1(\xi)=K_0(\xi)+\xi^{-1}K_1(\xi)$ in combination with the
asymptotic expressions $K_0(\xi)=-\ln\xi+\OO(1)$, $K_1(\xi)=
\xi^{-1}+ \OO(\ln\xi)$ for $\xi\to 0$, we find that
\begin{equation}
\left\| f'_{R,\kappa} \right\|^2_{L^2(\R^2)} < -\,{C \over
\ln(\kappa R) }
\end{equation}
holds for a positive constant $C$ and $\kappa R$ small enough.
This makes it possible to estimate the first term at the \rhs of
(\ref{energy form}) using the fact that the functions $\chi_j$ are
bounded outside $\BB_R$; recall that $\chi_j(x)= \OO(|x|^{-F+j})$
at large distances and $F-j \ge 0$.

We will show that $\int_{\Sigma} B(x)|\sum_j\alpha_j\chi_j(x)|^2\, d^2x >
0$ also holds in this situation again by assuming the opposite.
Indeed, let us set $h_j:=h\chi_j$ with a real-valued $h\in C_0^2(\Sigma)$
in (\ref{psi}); then the fourth term at the \rhs of
(\ref{energy form}) acquires the form
$$ 2\varepsilon\, \int_{\Sigma} \left|\, \sum_{j=0}^{[F]} \alpha_j
\chi_j(x) \right|^2 h(x) B(x)\, d^2x\,. $$
Since $B$ is nonzero, it is possible to choose $h$ so that the
integral is strictly negative. Taking $\varepsilon$ positive
and small enough the sum of the last four terms
at the \rhs of (\ref{energy form}) with $\mu=2$ can be made negative,
since the the linear term (in $\varepsilon$) prevails over
the quadratic ones and the third term is supposedly nonpositive.
The first term,
$$ \int_{\R^2} \left|f'_{R,\kappa}(r)\right|^2 \left|\,
\sum_{j=0}^{[F]} \alpha_j \chi_j(x) \right|^2 d^2x\,, $$
is positive, but the $\chi_j$'s are bounded outside $\BB_R$ and
$|\alpha|=1$, so it can be made sufficiently small by a suitable
choice of $\kappa$. Using again the supersymmetry property,
$D^*D+2B=DD^*$, we arrive at the absurd conclusion that
$\|D^*\psi_{\alpha}\|^2 < 0$.

Hence we can take finally the trial functions (\ref{psi}) with
$f_{R,\kappa}$ given by (\ref{tails}) and $\varepsilon=0$,
which yields the estimate
\begin{equation} \label{non L^2 bound}
(\psi_{\alpha},H_P^{(-)}(A)\psi_{\alpha})< -\,{C\over \ln(\kappa
R)}\, \max_{0\le j\le [F]} \|\chi_j\|^2_{\infty}- {1\over
2}\,(g-2)\,\min_{|\alpha|=1}\,\int_{\Sigma} B(x)|\psi_{\alpha}(x)|^2\, d^2x\,.
\end{equation}
The second term at the \rhs is strictly negative if $g>2$,
since $\int_{\Sigma}B(x)\,|\psi_{\alpha}(x)|^2\,d^2x > 0$ for
any $\alpha$ in a compact set (surface of the hypersphere $|\alpha|=1$),
and it dominates the sum for $\kappa$ small enough.

To conclude the proof of the first claim, one has to check
that the trial functions (\ref{psi}) span indeed a subspace of
dimension $1+[F]$. This follows readily from the linear
independence of $\psi_j:= f_{R,\kappa}\chi_j,\,
j=0,\dots,[F]$; recall that the $\chi_j$'s are linearly independent and
coincide with $\psi_j$ at least in the set $\BB_R$.

If $F=0$, the function $\widetilde\chi_0(x) := e^{\phi(x)}$
which solves $D^*\widetilde\chi_0=0$ is also bounded at large
distances and we can apply the analogous argument to the operator
$H_P^{(+)}(A) = DD^* + {1\over 2}\,(g-2)B$. Using a properly
chosen function $\widetilde\psi_0= f_{R,\kappa}\widetilde\chi_0
+\varepsilon h$, we can show that $\int_{\Sigma}
B(x)|\widetilde\chi_0(x)|^2\, d^2x < 0$, so $(\widetilde\psi_0,
H_P^{(+)}(A)\widetilde\psi_0) <0$ for $g>2$ and small $\kappa$; hence
$H_P^{(+)}(A)$ has a bound state as well. \quad
$\Box$
 \vspace{3mm}

\noindent {\bf Remarks.} (a) The argument fails only if $B=0$,
since then $\chi_0(x)= \widetilde\chi_0(x)= 1$ and the matrix
elements $\langle B\rangle_{\chi_0}$ and $\langle
B\rangle_{\widetilde\chi_0}$ are zero. \\ [1mm]
(b) Instead of the tail modification (\ref{tails}) a simpler one
--- $f_R(r):= f(r/R)$, where $f\in C_0^{\infty}(\R_+)$ is such
that $f(u)=0$ for $u\ge 2$ --- was employed in \cite{BEZ2}. The
kinetic energy is in this case estimated by
$$ {1\over R^2} \int_{\R^2} \left|f'\left(r\over R \right)
\right|^2\,  \left|\, \sum_{j=0}^{[F]} \alpha_j \chi_j(x)
\right|^2\, d^2x\le C\|f'\|^2_{\infty} R^{-2(F-[F])} $$
for a positive $C$. It is clear that one can handle in this way
the whole problem except for the case $F$ integer.

%%%%%%%%%%%%%%%%%%%%%%%%%%%%%%%%%%%%%%%%%%%%%%%%%%%%%%%%%%%%%%%%%%%%

\setcounter{equation}{0}
\section{Weakly bound states in two dimensions}

Schr\"odinger operators in dimension one and two can have bound
states for arbitrarily weak potentials, so the behaviour of the
ground state in these cases is of particular interest. The
corresponding asymptotic formulae, known already to Landau and
Lifshitz \cite{LL}, were analyzed rigorously in \cite{Si,BGS,Kl}.
If we make a digression to the subject here, it is because we want to
call attention to interesting aspects of the case when the
interaction is switched off in a nonlinear way.

It is sufficient, of course, to describe peculiarities of the
nonlinear case. Consider thus a two-dimensional Schr\"odinger
operator family
\begin{equation} \label{2D Hamiltonian}
H(\lambda) = -\Delta + V(\lambda,x)
\end{equation}
on $L^2(\R^2)$ with $ \lambda$ belonging to an interval
$[0,\lambda_0]$, where the potentials satisfy
\begin{equation} \label{potential expansion}
V(\lambda,x) = \lambda V_1(x) + \lambda^2 V_2(x) + W(\lambda,x)
\end{equation}
with
\begin{equation} \label{potential remainder}
|W(\lambda,x)| \le \lambda^3 V_3(x)
\end{equation}
and
\begin{equation} \label{potential regularity}
V_j \in L^{1+\delta}(\R^2) \cap L(\R^2, (1+|x|^{\delta})\,d^2x)\,,
\quad j= 1,2,3\,,
\end{equation}
for some $\delta>0$. By the Birman-Schwinger principle,
(\ref{2D Hamiltonian}) has an eigenvalue $\epsilon(\lambda)=
-\kappa^2$ iff the integral operator $K_{\kappa}$ with the
kernel
$$ K_{\kappa}^{\lambda}(x,y) = |V(\lambda,x)|^{1/2}\,R_0(\kappa;x,y)
\,V(\lambda,y)^{1/2} $$
(where $V^{1/2}:=|V|^{1/2}\,{\rm sign\,}V$) has an eigenvalue
$\,-1\,$; here
\begin{equation} \label{free kernel}
R_0(\kappa; x,y) = {1\over 2\pi}\, K_0(\kappa|x-y|)
\end{equation}
is the kernel of the free resolvent $(-\Delta +\kappa^2)^{-1}$.
A standard trick is then to split the operator under consideration
into two parts, $K_{\kappa}^{\lambda} = L_{\kappa}^{\lambda} +
M_{\kappa}^{\lambda}$, where the former is rank-one with the
kernel $L_{\kappa}^{\lambda}(x,y) =-{1\over 2\pi}\,
|V(\lambda,x)|^{1/2}\, \ln\kappa\, V(\lambda,y)^{1/2}$, while the
latter is regular as $\kappa\to 0+$ and, in this limit, has
kernel
$$ M_0^{\lambda}(x,y) = -\,{1\over 2\pi}\, |V(\lambda,x)|^{1/2}
\left\{ \gamma+  \ln{|x-y|\over 2} \right\}
V(\lambda,y)^{1/2}\,, $$
where $\gamma$ is the Euler constant. Now we employ the
identity
$$ (I+K_{\kappa}^{\lambda})^{-1} = \left[I+
(I+M_{\kappa}^{\lambda})^{-1} L_{\kappa}^{\lambda} \right]^{-1}
(I+M_{\kappa}^{\lambda})^{-1}\,, $$
where the existence of the inverses at the \rhs for sufficiently
small $\lambda$ follows from the assumptions made about the
potential in the same way as in \cite{Si}. The spectral problem is
thus reduced to finding a singularity of the square bracket, which
leads to an implicit equation. If we put $u:=
(\ln\kappa)^{-1}$, it can be written as
\begin{equation} \label{implicit}
u - {1\over 2\pi}\int V(\lambda,x)^{1/2}\,
(I+M_{\kappa}^{\lambda})^{-1}(x,y)\,|V(\lambda,y)|^{1/2}\,d^2x\,d^2y
=0\,,
\end{equation}
and used to derive the Taylor expansion of the function
$\lambda\mapsto u(\lambda)$; a weakly bound state with the
eigenvalue $\epsilon(\lambda)= -e^{2/u(\lambda)}$ exists iff
$u(\lambda)<0$ around the origin.

In the linear case, $V=\lambda V_1$, we get from here the
usual expansion
\begin{equation} \label{linear}
u(\lambda) = {\lambda\over 2\pi}\int V_1(x)\,d^2x +
{\lambda^2\over 4\pi^2}\int V_1(x) \left\{
\gamma+  \ln{|x-y|\over 2} \right\} V_1(y)\, d^2x\, d^2y +
\OO(\lambda^3)\,,
\end{equation}
which shows that a bound state exists iff $\int
V_1(x)\,d^2x \le 0$ (the second term is negative if the
potential $V_1$ is nontrivial and has zero mean \cite{Si}).

For a potential family (\ref{potential expansion}) nonlinear in
$\lambda$ the sign of $\int V_1(x)\,d^2x$ is again
decisive. An interesting situation arises, however, if the linear
part has zero mean,
\begin{equation} \label{zero mean}
\int V_1(x)\,d^2x = 0\,.
\end{equation}
Replacing $\lambda V_1$ with (\ref{potential expansion})
in (\ref{implicit}) and expanding in powers of $\lambda$
we find that
\begin{equation} \label{nonlinear}
u(\lambda) = \lambda^2 \left\{{1\over 2\pi}\int
V_2(x)\,d^2x + {1\over 4\pi^2}\int V_1(x)\,
\ln|x-y|\, V_1(y)\, d^2x\, d^2y \right\} + \OO(\lambda^3)
\end{equation}
holds in this case (the term with $\gamma- \ln 2$
splits into a product of  one-dimensional integrals and vanishes
too). We arrive at the following conclusion. \vspace{3mm}

\noindent {\bf Proposition 1.} An operator family (\ref{2D
Hamiltonian}) with the potential satisfying (\ref{potential
expansion})--(\ref{potential regularity}) and (\ref{zero mean}) has
a weakly bound state provided the leading coefficient in
(\ref{nonlinear}) is negative. If it is positive, no bound state
exists for small $\lambda$. \vspace{2mm}

\noindent The formula (\ref{nonlinear}) yields also the asymptotic
behaviour of the corresponding eigenvalue $\epsilon(\lambda)=
-e^{2/u(\lambda)}$. We will not inquire about the critical case
when the second-order coefficient also vanishes.

%%%%%%%%%%%%%%%%%%%%%%%%%%%%%%%%%%%%%%%%%%%%%%%%%%%%%%%%%%%%%%%%%%%%%

\setcounter{equation}{0}
\section{The centrally symmetric case}

Let us return now to the Pauli operator (\ref{Pauli}) and consider
the situation when the field is centrally symmetric, so the vector
potential can be chosen in the symmetric gauge, $\vec{A}(x)=
\lambda A(r) \vec e_{\varphi}\,$, with $A(r)=r^{-1}\int_0^rB(r')\,r'\,dr'$.
We have introduced the positive parameter $\lambda$ in order
to discuss how the spectral properties depend on the field
strength. We can perform a partial-wave decomposition and replace
(\ref{Pauli}) by the family of operators
\begin{equation} \label{partial wave Hamiltonian}
H_{\ell}^{(\pm)}(\lambda) =-\,{d^2\over dr^2} -{1\over r}\,
{d\over dr} + V_{\ell}^{(\pm)}(\lambda,r)\,, \qquad
V_{\ell}^{(\pm)}(\lambda,r):= \left(\lambda A(r)+
{\ell\over r} \right)^2 \pm {\lambda\over 2}\,g B(r)
\end{equation}
on $L^2(\R^+,r\,dr)$. In \cite{BEZ1} these operators were used
to discuss the behaviour of an electron in the magnetic field
induced by a localized rotating electric current. We need not
insist on that here, assuming only that the field is locally
integrable with $B(r)= \OO(r^{-2-\delta})$ as $r\to\infty$. However,
we will be interested primarily in the situation typical for
current-induced magnetic fields, in which the field has zero mean
(\ie, $F=0$) since the flux lines are closed in $\R^3$.

It is shown in \cite{BEZ1} under stronger assumptions ---
involving a smoothness and a faster decay  of the field --- that
each orbital Hamiltonian $H_{\ell}^{(-)}(\lambda)$ has a bound
state for $\lambda$ large enough, the critical values for
emergence of these states being, of course, $\ell$-dependent. This
result relies only on the behaviour of $V_{\ell}^{(-)}(\lambda,r)$
around the origin and is thus independent of the fact that $F=0$,
the important point being that $g>2$ so the ground-state energy of
the harmonic oscillator obtained in the limit $\lambda\to\infty$
is negative
--- \cf\cite{BEZ1}.

The ``spin-up'' Pauli operator $H_P^{(+)}(\lambda)$ may exhibit a
less intuitive behaviour as suggested by Theorem~1. If $F=0$ for a
{\em compactly supported} field, then $H_P^{(+)}(\lambda)$ has
also a bound state for any $\lambda>0$. Recall that Theorem~1 says
nothing about the size of $\Sigma$, it may be quite large.
Inspecting the shape of the effective potentials
$V_{\ell}^{(\pm)}(\lambda,r)$ for the two cases we see that the
states with different spin orientations are supported in different
regions: ``spin-down'' states in the vicinity of the origin (out
of the centrifugal barrier for $\ell\ne 0$), while the ``spin-up''
state at large distances where (for an arbitrary but fixed
$\lambda$) the magnetic field term dominates slightly over the
quadratic one in $V_0^{(+)}(\lambda,r)$ creating a shallow
potential well.

Let us examine the {\em weak-coupling behaviour} in the case of a
vanishing total flux, $F=0$, in the field with a tail, $B(r)=
\OO(r^{-2-\delta})$; no smoothness assumption is made. If $\ell\ne
0$, the first term in $V_{\ell}^{(\pm)}(\lambda,r)$ is bounded
below by $\lambda v(r)$ for a suitably chosen positive function
$v$ of compact support (the simplest choice is $v(r) =
c\,\Theta(r_0-r)$ for appropriate $c$ and $r_0$). Since the second
term does not contribute to $\int_0^{\infty}
V_{\ell}^{(\pm)}(\lambda,r)\,r\,dr$ which determines the linear
part of the weak--coupling behaviour, it follows from
(\ref{linear}) and the minimax principle that the discrete
spectrum of $H_{\ell}^{(\pm)}(\lambda)$ is empty for $\lambda$
small enough, when the centrifugal barrier prevents binding.

The interesting case is, of course, the s-wave part, where the
effective potential acquires the form (\ref{potential expansion})
with the last term absent and
\begin{equation} \label{potential coefficients}
V_1(r)= \,\pm\,{g\over 2}\,B(r)\,, \qquad V_2(r)= A(r)^2\,.
\end{equation}
In view of the assumption about the field, $r \mapsto A(r)$
is absolutely continuous and
$\OO(r^{-1-\delta})$, so the condition (\ref{potential
regularity}) is satisfied. It remains to evaluate the second
integral in (\ref{nonlinear}): we have
\begin{eqnarray*}
\lefteqn{{1\over 4\pi^2}\,\int_{\R^2\times\R^2} V_1(x)\,
\ln|x-y|\, V_1(y)\, d^2x\, d^2y} \\ %&& \\
&& =\, {g^2\over 4}\:
{1\over 2\pi}\, \int_0^{\infty} dr\, rB(r)\, \int_0^{\infty} dr'\,
r'B(r')\, \int_0^{2\pi}\, \ln\left[\, r^2\! +r'^2\! -2rr'
\cos\varphi\, \right]^{1/2}   d\varphi\,.
\end{eqnarray*}
By \cite[4.224.9]{GR} the last integral equals $ 2\pi
\ln\max(r,r')$; we substitute this into the formula and
integrate repeatedly by parts using $rB(r)= (rA(r))'$.
This yields
$$ -\, {g^2\over 4}\, \int_0^{\infty} \left( \int_r^{\infty}
A(r')\, dr' \right) B(r)\, r\,dr =  -\, {g^2\over 4}\,
\int_0^{\infty} dr\, A(r)^2\, r\,dr\,. $$
We arrive thus at the following conclusion. \vspace{3mm}

\noindent {\bf Proposition 2.} Let a spherically symmetric
magnetic field $B$ be locally integrable with
$B(r)= \OO(r^{-2-\delta})$ and vanishing flux,
$F=0$. Then each of the operators $H_0^{(\pm)}(\lambda)$
with $g>2$ has a negative eigenvalue for $\lambda$ small
enough. \vspace{3mm}

\noindent {\bf Remarks.} (a) The relation (\ref{nonlinear}) yields
also the asymptotic behaviour of the bound state energy,
\begin{equation} \label{weak asympt}
\epsilon^{(\pm)}(\lambda) \approx - \exp\left\{ -\left(
{\lambda^2\over 8}\, (g^2-4)\, \int_{0}^{\infty} A(r)^2\,r\, dr
\right)^{-1} \right\}
\end{equation}
as $\lambda\to 0$ with the usual meaning of $\approx$
(\cf\cite{Si}). The leading term is thus the same for both spin
orientations; however, since $g\ne 2$, the second theorem of
\cite{AC} does not apply and the degeneracy may be lifted in the
next order. \\ [1mm]
(b) Notice that the argument of the previous section cannot be
applied to compactly supported fields with a nonzero flux, since
the corresponding vector potential has then a too slow decay,
$A(r)= \OO(r^{-1})$, and consequently $V_1\not\in L(\R^2,
(1+|x|^{\delta})d^2x)$. One may ask whether the asymptotics is
nevertheless $\epsilon(\lambda)\approx \exp \left(-\, {4\over
\lambda Fg}\right)$ as it follows from a formal application of
(\ref{linear}). The example worked out in \cite{CFC} leads to the
conclusion that it is not the case --- see Eq.\ (17) of that paper.
The question about the asymptotic behaviour thus remains open. \\
[1mm]
(c) Another open question is whether the bound state of
$H_0^{(\pm)}(\lambda)$ survives generally for $\lambda$ large if
the field is not compactly supported.

%%%%%%%%%%%%%%%%%%%%%%%%%%%%%%%%%%%%%%%%%%%%%%%%%%%%%%%%%%%%%%%%%%%%%

\setcounter{equation}{0}
\section{Non-symmetric weak coupling revisited}

By different means, the result of the previous section complements
the zero-flux part of Theorem~1 in the weak-coupling case. While
imposing the symmetry requirement, it relaxes the assumptions on
the field decay. Here we want to show that the above argument can
be carried through for non-symmetric fields as well under mild
regularity assumptions; the price we shall pay is to have a weaker
form of the asymptotic formula (\ref{weak asympt}) only. Specifically,
suppose that
\begin{equation} \label{decay}
|B(x)|\le C_1 \langle x\rangle^{-2-\delta}\,, \qquad \int_{\R^2}
{|B(y)| \over |x-y|}\, d^2y \le C_2 \langle
x\rangle^{-1-\delta}\,,
\end{equation}
where $\langle x\rangle:= \sqrt{1+r^2}$. Then we have the
following result. \vspace{3mm}

\noindent {\bf Proposition 3.} Let a magnetic field $B$ satisfy
the conditions (\ref{decay}) for some $C_1, C_2,\delta>0$, and let
$F=0$. Then each of the operators $H_P^{(\pm)}(\lambda)$ with
$g>2$ has a negative eigenvalue for $\lambda$ small enough.
\vspace{2mm}

\noindent {\em Proof$\,$} is based on two observations. The first
one concerns the ``mixed'' term $2iA\cdot\nabla$ in the
Hamiltonian; we shall show that it does not contribute to the
energy form for real-valued functions (the other ``mixed'' term,
$i\nabla\cdot A$, vanishes in the
gauge we have been adopting). More specifically, take a
real-valued $\psi\in C_0^2(\R^2)$, \ie, twice differentiable
with compact support. For the sake of brevity, we write the
vector potential components as $A_i= -\epsilon_{ij}\partial_j
\phi$, where $\epsilon$ is the two-dimensional Levi-Civita
tensor, and employ the convention of summation over repeated
indices; then
\begin{eqnarray*}
(\psi,A\cdot\nabla\psi) &\!=\!& - \int_{\R^2} \psi(x)\,
\epsilon_{ij}\, (\partial_j\phi)(x)\, (\partial_i\psi)(x)\, d^2x \\
\\
&\!=\!& -\,{1\over 2}\, \lim_{R\to\infty} \int_{\BB_R}
\epsilon_{ij}\, (\partial_j\phi)(x)\, (\partial_i\psi^2)(x)\, d^2x \\
\\
&\!=\!& -\,{1\over 2}\, \lim_{R\to\infty} \int_{\BB_R}
\epsilon_{ij} \left\{ (\partial_j(\phi\,\partial_i \psi^2))(x) -
\phi(x)\, (\partial_i \partial_j \psi^2) \right\}\, d^2x \\
\\
&\!=\!& {1\over 2}\, \lim_{R\to\infty} \oint_{\partial\BB_R}
\phi(x)\, (\nabla\psi^2)(x)\cdot d\vec\ell(x) \,=\, 0\,.
\end{eqnarray*}
The third line is obtained from the second
using integration by parts. Its second term vanishes because
$\partial_j \partial_i \psi^2$ is symmetric with respect to
the interchange of indices and is contracted with the
anti-symmetric symbol $\epsilon_{ij}$. The remaining term is
rewritten by means of
the Stokes theorem and vanishes in the limit since
$\nabla\psi^2$ has compact support.

The second observation is that the relation between the two
integrals which we found in the proof of Proposition~2 by
explicit computation in polar coordinates is valid generally. To
see this, let us rewrite $\int A(x)^2\,d^2x$ by means of the
first Green identity:
\begin{eqnarray}
\int_{\R^2} A(x)^2 d^2x &\!=\!& \lim_{R\to\infty} \int_{\BB_R}
(\nabla\phi(x))^2\, d^2x \nonumber \\
\nonumber \\
&\!=\!&\lim_{R\to\infty} \int_{\BB_R}\left\{(\nabla\cdot(\phi\nabla\phi))(x)
- \phi(x)\, (\nabla^2\phi)(x) \right\} d^2x \label{ns identity} \\
\nonumber \\
&\!=\!& \lim_{R\to\infty} \oint_{\partial \BB_R}
\phi(x)\,(\nabla\phi)(x)\cdot d\vec\sigma(x) - \lim_{R\to\infty}
\int_{\BB_R} \phi(x)B(x)\, d^2x\,; \nonumber
\end{eqnarray}
in the second integral we have used $\triangle\phi=B$
and the first one was rewritten by means of Gauss theorem. Our
aim is now to use the conditions (\ref{decay}) to demonstrate that
the first integral at the \rhs vanishes in the limit. The decay
hypothesis about the field yields
\begin{eqnarray*}
|\phi(x)| &\!\le\!& {1\over 2\pi}\, \int_{|y-x|\le 1} |B(y)|\,
|\ln|x-y||\, d^2y + {1\over 2\pi}\, \int_{|y-x|\ge 1} |B(y)|\,
\ln|x-y|\, d^2y \\
\\
&\!\le\!& {C_1\over 2\pi}\, \int_{|z|\le 1}
|\ln |z||\, d^2z + {C_1\over 2\pi}\, \int_{|z|\ge 1} \langle x-z
\rangle^{-2-\delta}\, \ln|z|\, d^2z\,.
\end{eqnarray*}
Denote $\ln_{+} u:= \max(0,\ln u)$. Then for any $\eta>0$
there is a $K_{\eta}>0$ such that
$$ \ln_{+}|z| < K_{\eta}\, \langle x-z \rangle^{\eta}\,
(1+\ln_{+}|x|) \;; $$
this follows from the fact that
$$ \limsup_{\zeta,\xi\to\infty} {\ln_{+}\zeta \over
\left(1+|\xi-\zeta|^2\right)^{\eta/2} (1+\ln_{+}\xi)} \le 1 $$
in the first quadrant, and the function under the limit is
continuous there. The second one of the above integrals is thus
estimated by
$$ {C_1K_{\eta}\over 2\pi}\,(1+\ln_{+}|x|)\, \int_{\R^2} \langle y
\rangle^{-2-\delta+\eta}\, d^2y\,. $$
For $\eta < \delta$ the last integral is convergent; hence
there is a $C'>0$ such that $|\phi(x)| \le C' \ln|x|$ as
$|x|\to\infty$. The second one of the conditions (\ref{decay})
implies
$$ |A(x)| = |(\nabla\phi)(x)| \le {C_2 \over 2\pi}
\langle x\rangle^{-1-\delta}\,, $$
so the first integral at the \rhs of (\ref{ns identity}) is
$o(R^{-\delta'})$ for any $\delta' < \delta$ and vanishes in the
limit that we have set out to prove. Substituting (\ref{phi}) for
$\phi$ in the second one we arrive finally at the identity
\begin{equation} \label{AB identity}
\int_{\R^2} A(x)^2\, d^2x = -\,{1\over 2\pi}\,\int_{\R^2\times\R^2}
B(x)\, \ln|x-y|\, B(y)\, d^2x\, d^2y\;;
\end{equation}
the conditions (\ref{decay}) ensure that both integrals exist.
Now it is easy to conclude the proof. We have
\begin{eqnarray} \label{inf sigma}
\inf\sigma\left(H_P^{(\pm)}(\lambda)\right)  &\!=\!& \inf\left\{\,
\left(\psi, H_P^{(\pm)}(\lambda) \psi\right):\; \psi\in D\left(
H_P^{(\pm)}(\lambda) \right)\, \right\} \nonumber \\
\nonumber \\
&\!\le \!&\inf\left\{\, \left(\psi, H_P^{(\pm)}(\lambda) \psi\right):\;
\psi\in C_0^2(\R^2)\,,\: \psi=\bar\psi\, \right\} \nonumber \\
\nonumber \\
&\!=\!&\inf\sigma\left(\widetilde H_P^{(\pm)}(\lambda)\right),
\end{eqnarray}
where
$$ \widetilde H_P^{(\pm)}(\lambda) := -\Delta +\lambda^2 A(x)^2 \pm
{\lambda g\over 2} B(x)\,. $$
The last equality in (\ref{inf sigma}) is due to the fact that
$C_0^2(\R^2)$ is a core of $\widetilde H_P^{(\pm)}(\lambda)$. It
is now sufficient to apply Proposition~1 to the operator $\widetilde
H_P^{(\pm)}(\lambda)$ and to employ the identity (\ref{AB identity}).
\quad $\Box$
\vspace{3mm}

\noindent {\bf Remark.} In view of the estimate used in the proof,
the relation (\ref{weak asympt}) is now replaced by the asymptotic
inequality
\begin{equation} \label{ns weak asympt}
\epsilon^{(\pm)}(\lambda) \,\aleq\, -\, \exp\left\{ -\left(
{\lambda^2\over 16\pi}\, (g^2-4)\, \int_{\R^2} A(x)^2\,d^2x
\right)^{-1} \right\};
\end{equation}
the question whether the \rhs is still the lower bound remains
open.

%%%%%%%%%%%%%%%%%%%%%%%%%%%%%%%%%%%%%%%%%%%%%%%%%%%%%%%%%%%%%%%%%%%%%%%%%%%%

\subsection*{Acknowledgments}

The authors thank for the hospitality extended to them at the
institutes where parts of the work were done: P.E.\ to CPT Luminy
and V.Z.\ to NPI \v{R}e\v{z}. The research has been partially
supported by GACR and GAAS under the contracts 202/96/0218 and
1048801. R.M.C.\ was supported by CNPq and NSF under grant
No.~PHY94-07194.

%%%%%%%%%%%%%%%%%%%%%%%%%%%%%%%%%%%%%%%%%%%%%%%%%%%%%%%%%%%%%%%%%%%%%%%%%

\end{document}